\documentclass[a4paper,twoside]{article}
\usepackage{amsmath}
\usepackage{amsthm}
\usepackage{fullpage}
\usepackage{physics}
\usepackage{epsfig}
\usepackage[linesnumbered,ruled,vlined]{algorithm2e}
\usepackage{color}
\usepackage{url}
\usepackage{subcaption}
\usepackage{calc}
\usepackage{amssymb}
\usepackage{amstext}
\usepackage{multicol}
\usepackage{pslatex}
\usepackage{apalike}
\usepackage{listings}
\usepackage[bottom]{footmisc}
\usepackage{booktabs}
\usepackage{multirow}
\usepackage{caption}
\usepackage{graphicx} 
\usepackage{lipsum}
\usepackage[table,xcdraw]{xcolor}
\usepackage{adjustbox}
\usepackage{SCITEPRESS}
\begin{document}

\twocolumn

\title{Differential Evolution Algorithm based Hyper-Parameters Selection of Convolutional Neural Network for Speech Command Recognition}

\author{\authorname{Sandipan Dhar\sup{1}\orcidAuthor{0000-0002-3606-6664},
Anuvab Sen\sup{2}\orcidAuthor{0009-0001-8688-8287}, Aritra Bandyopadhyay\sup{3}\orcidAuthor{0009-0003-5582-2431}, Nanda Dulal Jana\sup{1}\orcidAuthor{0000-0003-0631-9912},  Arjun Ghosh\sup{1}\orcidAuthor{0000-0003-1086-944X}, Zahra Sarayloo\sup{4}\orcidAuthor{0000-0001-9918-3625}} 
\affiliation{\sup{1}Computer Science and Engineering, National Institute of Technology, Durgapur, West Bengal, India}
\affiliation{\sup{2}Electronics and Telecommunication, Indian Institute of Engineering Science and Technology, Shibpur, Howrah, India}
\affiliation{\sup{3}Computer Science and Technology, Indian Institute of Engineering
Science and Technology, Shibpur, Howrah, India}
\affiliation{\sup{4}School of Computer Science, University of Waterloo, Ontario, Canada}
\email{sd.19cs1101@phd.nitdgp.ac.in, sen.anuvab@gmail.com, aritraxban@gmail.com,  rjun.cse@gmail.com, nandadulal@cse.nitdgp.ac.in, zsaraylo@uwaterloo.ca}
}

\keywords{Differential Evolution Algorithm, Genetic Algorithm, Convolutional Neural Network, Hyper-parameters Selection, Meta-heuristics, Speech Command Recognition, Deep Learning}

\abstract{Speech Command Recognition (SCR), which deals with identification of short uttered speech commands, is crucial for various applications, including IoT devices and assistive technology. Despite the promise shown by Convolutional Neural Networks (CNNs) in SCR tasks, their efficacy relies heavily on hyperparameter selection, which is typically laborious and time-consuming when done manually. This paper introduces a hyperparameter selection method for CNNs based on the Differential Evolution (DE) algorithm, aiming to enhance performance in SCR tasks. Training and testing with the Google Speech Command (GSC) dataset, the proposed approach showed effectiveness in classifying speech commands. Moreover, a comparative analysis with Genetic Algorithm-based selections and other deep CNN (DCNN) models highlighted the efficiency of the proposed DE algorithm in hyperparameter selection for CNNs in SCR tasks.}
\onecolumn \maketitle \normalsize \setcounter{footnote}{0} \vfill





\section{\uppercase{Introduction}}
Speech Command Recognition (SCR) is a subfield of Automatic Speech Recognition (ASR) focused on converting short spoken words into text \cite{SCC}. It's widely used in Internet of Things (IoT)-based smart home assistants, command-controlled wheelchairs for blind and disabled people, and AI-driven vehicles \cite{Survey_on_SCR}. Early SCR systems primarily used Hidden Markov Models (HMMs) \cite{HMM}, Gaussian Mixture Models (GMMs) \cite{GMM}, and Multi-Layered Perceptron models (MLPs) \cite{MLP}. Later, Recurrent Neural Networks (RNNs) \cite{RNN} and Long Short-Term Memory networks (LSTMs) \cite{LSTM} yielded significant improvements. However, Convolutional Neural Networks (CNNs), effective in handling 2D data dependencies, emerged as superior alternatives \cite{Survey_on_SCR}. Various input features have been considered while dealing with SCR tasks, like using a Depth-Wise Separable CNN (DS-CNN) for keyword recognition with mel-frequency spectral coefficients (MFSS) as input feature \cite{Depth-wise-seperable-CNN}, using mel-frequency cepstral coefficients (MFCC) as input for deploying a CNN for wheelchair control using speech commands \cite{Bakouri}, smoothed-spectrogram, mel-spectrogram, and cochleagram as input features for CNN-based voice command detection \cite{Cochleagram}. Kubanek et al. proposed a new approach where MFCC, time and spectrum are combined to be used as speech features for the recognition of speech commands using DCNN model \cite{Speech_Coding}. However, the performance of all these CNNs is highly dependent on selection of several crucial hyper-parameters.
\par
CNN models have hyper-parameters like number and type of convolution layers, filter count and size, pooling type, and activation function, which significantly influence performance in classification tasks, including SCR. Typically, hyper-parameters are manually selected based on experience, a process that is both time-consuming and tedious. Therefore, it becomes difficult to obtain the optimal configuration of a CNN model within a reasonable cost \cite{nas,enas,Arjun,Arjun1}. The paper employs the Differential Evolution (DE) algorithm \cite{devo} \cite{sen2023differential} \cite{sen2023comparative} \cite{mazumder2023benchmarking} \cite{sen2023differential2} to optimize CNN hyper-parameters for SCR tasks. Each individual in the DE algorithm represents a viable CNN architecture, with optimal hyper-parameters determined through standard DE operations like mutation, crossover, and selection. Spectrograms are used as input speech features for the CNN model. The dataset considered in this work is the Google Speech Command (GSC) dataset \cite{GSC}. The proposed DE algorithm-based hyper-parameters selection approach is compared with the Genetic Algorithm (GA) \cite{DE} based hyper-parameter selection, as well as with state-of-the-art deep CNN (DCNN) models namely ResNet-50, Inception-V3, Xception, VGG-16 and VGG-19 for SCR task. The work maintains a consistent basic CNN architecture (with a fixed number of convolution, pooling, and fully connected layers) for both DE and GA approaches while implementing automatic hyper-parameter selection. Experimental results demonstrate that the proposed method outperforms others, achieving higher accuracy.
\par
Rest of the paper is organized as follows. Sections 2 and 3 provide detailed overviews of the related work and preliminaries respectively. Section 4 includes the details of the dataset, training details and experimental setups. In Section 5, the proposed approach is briefly discussed. In Section 6, experimental results are presented and discussed. Finally, Section 7 concludes the paper and provides some aspects of the future research.
\section{\uppercase{Related Work}}

Hyperparameter optimization is a critical research area for achieving high-performance deep learning models. Techniques like Random Search, Grid Search, Bayesian Optimization \cite{Masum_Shahriar_Haddad_Faruk_Valero_Khan_Rahman_Adnan_Cuzzocrea_Wu_2021}, and Gradient-based Optimization \cite{maclaurin2015gradientbased} are used to find optimal hyperparameter configurations. Each method offers trade-offs in the computational efficiency, exploration of search space, and exploitation of discerned solutions. Genetic Algorithms were first utilized for modifying Convolutional Neural Network architectures in late 1900's, subsequently instigating a gamut of applications involving various nature-inspired algorithms in the domain of deep learning models. While many works compare evolutionary algorithms on computational models, no previous study comprehensively has applied evolutionary algorithms: Genetic Algorithm, Differential Evolution, across Convolutional Neural Networks architecture for isolated speech command recognition. These algorithms stand out due to their iterative population-based approaches, stochastic and global search implementation, and versatility in optimizing various problems. In this context, this paper aims to bridge the gap by conducting a comprehensive exploration of the application of nature-inspired and evolutionary algorithms, like Differential Evolution, in optimizing DCNN architectures for SCR. By delving into the intricacies of how these algorithms interact with lightweight CNN structures and comparing the performance in SCR with that of DCNN models, namely, VGG-16, VGG-19, Resnet-50, InceptionV3, Xception, this study aims to uncover a clearer understanding of their advantages and the limitations for the various other speech related tasks.
\section{\uppercase{Preliminaries}}

\subsection{Differential Evolution (DE)}

DE is a population-based optimization algorithm designed for non-linear, multi-modal optimization problems \cite{devo}. It iteratively refines a population of candidate solutions (individuals) through mutation, crossover, and selection operators, enhancing the individuals based on existing ones within the population. In order to apply DE, first a population size of $\textit{N}$ individuals is created, and each individual is represented by a d-dimensional vector $\mathbf{x}_{i}$ (where $i$ implies the $i^{th}$ individual). Thereafter, the population is randomly initialized within the search space. At each iteration, a new population with $\textit{N}$ individuals are generated by applying the following mutation, crossover and selection operators.
\par
\textbf{Mutation:} In mutation operation, distinct individuals from the population are selected. A widely used mutation scheme is $DE/rand/1$, where three distinct individuals from the population are randomly selected. Then, a mutant vector (also called donor vector) $\mathbf{v}^g_i$ is created as shown in Eq. (\ref{eq1}),
\begin{equation} \label{eq1}
 \mathbf{v}_{i}^{g}=\mathbf{x}_{r_1}^{g}+ F \times \left(\mathbf{x}_{r_2}^{g}-\mathbf{x}_{r_3}^{g}\right).  
\end{equation}
In Eq. (\ref{eq1}), $\mathbf{x}^{g}_{r_1}$, $\mathbf{x}^{g}_{r_2}$, and $\mathbf{x}^{g}_{r_3}$ are three distinct individuals (here, $g$ indicates generation). Whereas, ${r_1}$, ${r_2}$, and ${r_3}$ means randomly selected indices, and $F$ is the scaling factor that controls the magnitude of the mutation. 
\par
\textbf{Crossover:} In crossover operation, a trial vector $\mathbf{u}^{g}_{i}$ is generated by combining the donor vector  $\mathbf{v}^{g}_{i}$ and the original vector  $\mathbf{x}^{g}_{i}$ using crossover operation. The crossover operation (binomial crossover) is explained in details as follows, 
\begin{equation} \label{eq2}
  \mathbf{u}_{j,i}^{g}=\begin{cases}
    \mathbf{v}_{j,i}^{g} & if \hspace{0.11cm}{j}_{\text{rand}} (0,1) \leq {CR} \hspace{0.11cm} or\hspace{0.11cm} j = \hspace{0.11cm} \delta  \\
    \mathbf{x}_{j,i}^{g} & Otherwise. \\
  \end{cases}
\end{equation}
In this context, $\mathbf{{u}}_{j,i}^{g}$ represents the $j^{th}$ dimension of the $i^{th}$ individual at the $g^{th}$ generation. The crossover rate is denoted by $CR$, and $rand(0,1)$ represents a randomly generated number between 0 and 1. Additionally, $\delta$ refers to a random dimension $d$ selected from the range $(1,d)$ of $\mathbf{{u}}_{i}^{g}$.
\par
\hspace{1cm}

\textbf{Selection:}
In selection operation, the trial vector $\mathbf{u}^{g}_{i}$ is compared with the original vector $\mathbf{x}^{g}_{i}$. If the fitness of $\mathbf{u}^{g}_{i}$ is superior than $\mathbf{x}^{g}_{i}$, then replacement of $\mathbf{x}^{g}_{i}$ with $\mathbf{u}^{g}_{i}$ is carried out in the next generation. Otherwise, $\mathbf{x}^{g}_{i}$ is kept unchanged. The above three steps are repeated until a stopping criterion is met (the stopping criterion varies from problem to problem).

\section{\uppercase{EXPERIMENTAL DETAILS}}
\subsection{Dataset Description}

The proposed DE-based hyper-parameters selection approach is trained and tested on google speech command (GSC) dataset \cite{GSC}. In this work $8$ speech commands from GSC dataset are considered namely “down”, “go”, “left”, “no”, “right”, “stop”, “up”, “yes”. Here, total $8000$ speech samples are considered by taking $1000$ samples belonging to each speech commands. The dataset is split into training, validation and test set. In this work, the model is trained with $6400$ training samples and $1000$ validation samples. After the completion of training the trained model is tested with $600$ test samples. The time span of each audio sample considered is of 1 second or less and the sampling rate is 16kHz.
\subsection{Experimental Setups}
The experiments of this work are implemented in Python
3.10.11 using three libraries as Tensorflow 2.11.0, Tensorflow
built in Keras, and Numpy 1.22. The audible speech data
samples are preprocessed using Librosa 0.10.0. The experiments were performed in a Google Colaboratory environment using A100 GPU.
\section{\uppercase{Proposed Approach}}
This section explicitly describes the proposed DE algorithm-based hyper-parameter selection of CNN model for speech command recognition task. In the pre-processing phase, each speech sample is converted into mel-spectrogram \cite{Mel-Spectrogram} of shape $124\times129$, in order to make the input data compatible to work with $2D$ CNN. First, an overall framework of the proposed method is presented followed by the main components of the method. These include encoding scheme, population initialization, fitness evaluation, mutation, crossover, and selection operation of DE, concerning the optimal hyper-parameter selection for the CNN model.

\subsection{The Overall Framework}
\begin{figure*}
\centering
\includegraphics[height=6cm, width=15.4cm]{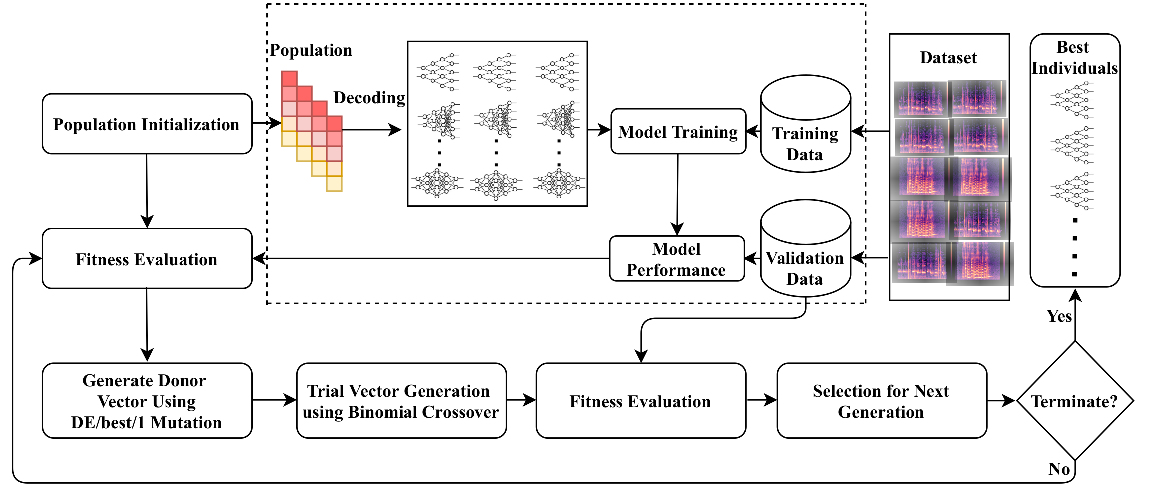}
\caption{The working mechanism of the DE algorithm based hyper-parameters selection approach for the SCR task.}  
\label{Fig-1}
\end{figure*}

The overall framework of the proposed approach is depicted in Fig. \ref{Fig-1}.

The DE algorithm starts with a population of $N$ individuals, each representing a CNN architecture which is trained on the training dataset ($D_{train}$) and evaluated for fitness on the validation dataset ($D_{valid}$) in terms of model accuracy. The associated hyper-parameters of CNN models are evolved through mutation and crossover operations of DE. These processes are repeated with a maximum number of generations. The optimal hyper-parameters of CNN architecture are selected from the best individual based on their fitness value and tested on the test dataset to determine the model's final performance.

\subsection{Encoding Scheme}

Designing an appropriate encoding process is a difficult task in any algorithm, as it determines how each individual is represented as a CNN structure. To address this, a standard layer-based encoding scheme is proposed in this work. This adopts the widely popular VGG-16 CNN model design \cite{VGG-16}. The VGG-16 model is composed of three types of layers - convolution, pooling, and fully connected (FC) arranged sequentially. Each individual's length is fixed with a total of 16 layers, following VGG-16 model. The hyper-parameters for each layer are determined based on pre-defined ranges for the purpose of designing and training a CNN model.

\subsection{Population Initialization}
The population in this context refers to the collection of individuals that are initially spread throughout the search space. The population is denoted as $\textit{P}$, consists of $\textit{N}$ individuals represented as $\textit{P}=\{\mathbf{x}_1, \mathbf{x}_2,\mathbf{x}_3,...,\mathbf{x}_N\}$. Every individual is regarded as a CNN model architecture with a fixed length similar to the VGG-16 model architecture. In addition, the corresponding hyper-parameters of the CNN model are initialized randomly within a set of pre-defined ranges defined in Table \ref{table1}.

\begin{table}[htbp]
\small 
\setlength{\tabcolsep}{6pt} 
\caption{Hyper-parameters and their ranges considered in the proposed work}
\centering
\begin{tabular}{ll}
\hline
\textbf{Hyper-parameters} & \textbf{Hyper-parameters range} \\
\hline
Convolution filter size & \{3$\times$3, 5$\times$5\} \\
Number of filters & \{16, 32, 64, 128, 256, 512\} \\
Activation function & \{`ReLU', `SELU', `ELU'\} \\
Optimization function & \{`SGD', `Adam', `Adagrad', \\ & `Adamax'\} \\
Drop-out rate & \{0.1, 0.2, 0.3, 0.4, 0.5\} \\
Number of neurons & \{128, 256, 512\} \\
\hline
\end{tabular}
\label{table1}
\end{table}

\begin{figure*}[ht]
\centering
\subfloat[Genotype]{%
  \includegraphics[height=1cm, width=7.4cm]{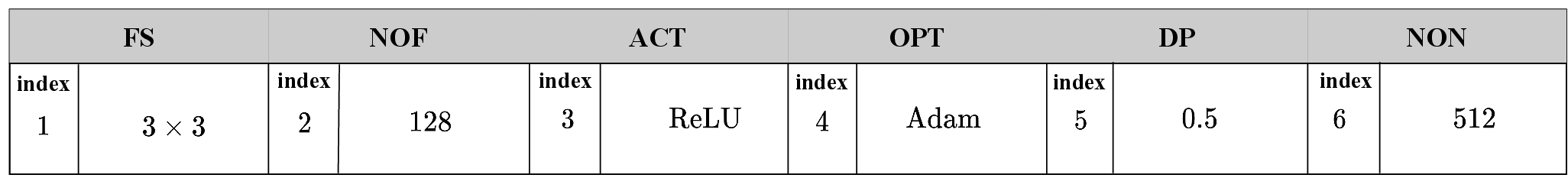}%
  \label{MCEP trajectories M2F}%
}
\hfill
\subfloat[Phenotype]{%
  \includegraphics[height=2cm, width=7.4cm]{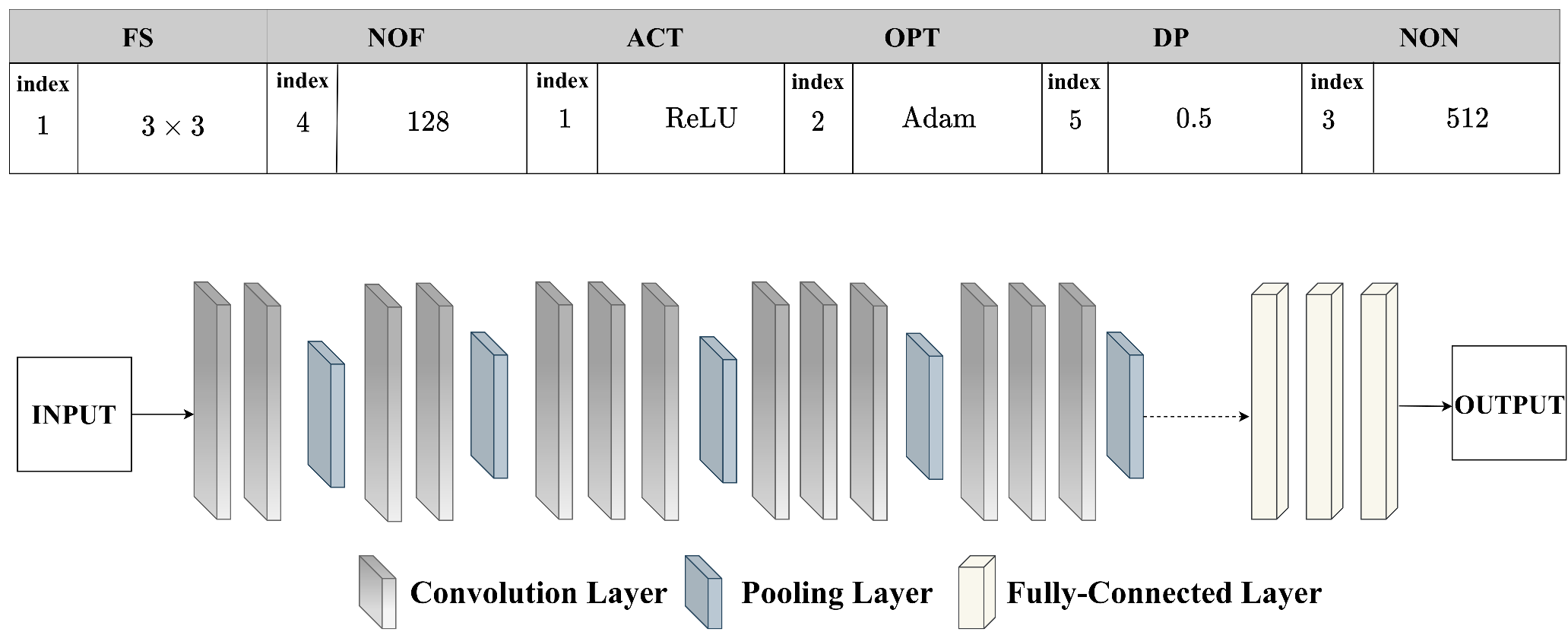}%
  \label{MCEP trajectories F2M}%
}
\caption{An example of genotype with its phenotype. The acronym used in genotype is: FS (Convolution Filter Size), NOF (Number of Filters), ACT (Activation function), OPT (Optimization function), DP (Drop-out rate), NON (Number of Neurons).}
\label{f1}
\end{figure*}

A layer-based approach is used to configure the hyper-parameter of each layer type, including convolution filter size, number of filters, activation function, optimizer, drop-out value and number of neurons in FC layers. In this work, the hyper-parameters of the pooling layer are considered as same as the VGG-16 model. Fig.\ref{f1} shows an example of a genotype along with its corresponding phenotype.

\subsection{Fitness Evaluation}\label{fitness}
In the proposed method, each individual in the population is evaluated based on their fitness. To calculate the fitness, every individual in the population $\textit{P}$ transforms itself into a CNN architecture and trains it with the training dataset $D_{train}$. The trained model is then evaluated on the validation dataset $D_{valid}$ using sparse categorical cross-entropy \cite{Cross_Entropy_Loss} as the fitness function due to its excellent performance in the SCR tasks.
\subsection{Mutation}

In DE, a mutant or donor vector is obtained by applying different mutant operations to the original vector of the current generation. In this study, the $DE/rand/1$ mutation scheme is used for simplicity and greater diversity in the hyper-parameters of CNN architecture at each generation. During the mutation phase, as described in Eq. (\ref{eq1}), a basic difference calculation is employed to compare the hyper-parameters of the chosen CNN model. In the proposed approach, two individuals ($\mathbf{x}_{r_2} \neq \mathbf{x}_{r_3}$) are selected randomly from the population \textit{P} which are different from the original vector $\mathbf{x}_{i}$. Then, the difference $(\mathbf{x}_{r_2}$-$\mathbf{x}_{r_3})$ is calculated based on the hyper-parameter values for each layer of CNN. After performing the difference calculation, the range of hyper-parameters for each layer is checked by boundary checking to ensure that they fall within specified limits. Next, the proposed approach selects another random individual, denoted as $\mathbf{x}_{r_1}$ and performs the computation with $(\mathbf{x}_{r_2}$-$\mathbf{x}_{r_3})$ to generate a donor vector (also called a mutant vector) $\mathbf{v}_{i}$ based on a scaling factor $F$. For this purpose, a random number $r[0,1]$ is generated for each dimension of $\mathbf{x}_{r_1}$. If ${r \leq F}$, the proposed method chooses a layer from $\mathbf{x}_{r_1}$.

\hspace{1.8cm}

Otherwise, it selects a layer from $(\mathbf{x}_{r_2}$-$\mathbf{x}_{r_3})$. Eq. (\ref{eq31}) specifies the mutation operation, where  $\mathbf{v}_{j,i}$ represents the $j^{th}$ dimension of the $i^{th}$ individual in the population $\textit{P}$.
\hspace{3.31cm}
\begin{equation} \label{eq31}
 \mathbf{v}_{j,i}=\begin{cases}
   \mathbf{x}_{j,{r_1}} & if\hspace{0.11cm} $r$\hspace{0.51cm}\leq\hspace{0.11cm} $F$ \\
    \lvert\mathbf{x}_{j, {r_2}}$-$\mathbf{x}_{j,{r_3}}\rvert & Otherwise \\
  \end{cases}
\end{equation}

\hspace{1.8cm}

Since we cannot calculate $(\mathbf{x}_{r_2}$-$\mathbf{x}_{r_3})$ for activation functions, as there is no defined ``difference" between them, we follow an encoding and rounding off strategy, encoding the activation functions with integers and then performing rounding off and boundary checking while decoding.
\subsection{Crossover}

To boost population diversity, a crossover operation follows the mutation operation in DE, exchanging components between the donor vector $\mathbf{v}_{j,i}$ and target vector $\mathbf{x}_{j,i}$ to form a new trial vector $\mathbf{u}_{i}$. Binomial crossover is employed, with the trial vector formation guided by crossover rate $\textit{CR}$ and a random number $\delta$. We defines $\delta$ value randomly one of the $j^{th}$ component of $\mathbf{v}_{i}$. Another random number $j_{rand}(0,1)$ is assigned for each dimension ($j$) of $\mathbf{u}_{i}$ that has the same length of $\mathbf{v}_{i}$. If the randomly generated number  $j_{rand}(0,1)$ is less than or equal to the crossover rate $CR$, or if $j$ is equal to $\delta$, then the $j^{th}$ value from the donor vector $\mathbf{v}_{i}$ is selected. Otherwise, the $j^{th}$ value is taken from the target vector $\mathbf{x}_{i}$. 
\hspace{1cm}

The proposed crossover operation is mathematically represented in Eq. (\ref{eq32}), where trial vector $\mathbf{u}_{j,i}$ represents the $j^{th}$ dimension of the $i^{th}$ individual for the target vector $\mathbf{x}_{i}$.
 \begin{equation} \label{eq32}
\mathbf{u}_{j,i}=\left\{\begin{array}{ll}
\mathbf{v}_{j,i} & \text { if } {j}_{\text {rand }}(0,1) \leq CR \hspace{0.11cm} or \hspace{0.11cm} {j}=\delta \\
\mathbf{x}_{j, i} & \text { Otherwise }
\end{array}\right.
\end{equation}
\subsection{Selection}

The selection stage chooses either the target vector $\mathbf{x}_{i}$ or trial vector $\mathbf{u}_{i}$ for the next generation based on their fitness values f, ensuring a constant population size across generations for stability. Each $\mathbf{x}_{i}$ in the population $\textit{P}$ is evaluated for its fitness, denoted as $f(\mathbf{x}_{i})$, using the fitness function. Also, the fitness of generated $\mathbf{u}_{i}$ is calculated using the same fitness function as for each $x_i$ and represented as $f(\mathbf{u}_{i})$. For the subsequent generation, i.e., $(g+1)$, the individual with higher fitness value is selected. Eq. (\ref{eq33}) mathematically presents the proposed selection strategy used in our proposed work.
\begin{equation} \label{eq33}
  \mathbf{x}_{i}^{g+1}=\begin{cases}
    \mathbf{u}_{i}^{g} & if \hspace{0.11cm} f(x_{i}^{g}) \leq \hspace{0.11cm} {f(u_{i}^{g})} \hspace{0.11cm}\\
    \mathbf{x}_{i}^{g} & Otherwise \\
\end{cases}
\end{equation}

A pseudocode implementation of the proposed Differential Evolution Algorithm is as follows:
\begin{algorithm}[htbp]
\renewcommand{\thealgocf}{}
\SetAlgoLined
\SetKwInOut{Input}{Input}
\SetKwInOut{Output}{Output}
\Input{$N = 15$, Dimension $D$, Scale Factor $F$, $CR$, Termination Criterion}
\Output{Best individual}

Initialize the population with $N$ random individuals in the search space;

\While{Termination Criterion is not met}{
\For{each individual $x_i$ in the population}{
Select three distinct individuals $x_{r_{1}}$, $x_{r_{2}}$, and $x_{r_{3}}$ from the population;

Generate a trial vector $v_i$ by mutating $x_{r_{1}}$, $x_{r_{2}}$, and $x_{r_{3}}$ using the differential weight $F$;
\begin{equation}
    v_{i} = x_{r_{1}} + F \times (x_{r_{2}}-x_{r_{3}})
\end{equation}

Perform crossover between $x_i$ and $v_i$ to produce a trial individual $u_i$ with the crossover probability $CR$;
\begin{equation}
u_{j, i} = 
\left\{
    \begin{array}{lr}
        v_{j, i}, & \text{if } p_{rand}(0, 1) \leq CR \\
        x_{j, i} & \text{else} 
    \end{array}
\right\}
\end{equation}

    \If{the fitness of $u_i$ is better than the fitness of $x_i$}{
        Replace $x_i$ with $u_i$ in the population\;
    }
}
}

\Return the best individual in the final population;

\caption{Differential Evolution}
\end{algorithm}

\section{\uppercase{RESULTS AND DISCUSSION}}

The parameters setting of the proposed work is based on the literature review of the conventional DE \cite{devo} and deep learning (DL) \cite{deep} implementations along with our limited computational resources. The population size and maximum generation are fixed at $10$ throughout the proposed algorithm\footnote{Code implementation of the proposed work is available at: \url{https://github.com/Techie5879/Hyperparameter-Optimization-CNN-Differential-Evolution}}. DE scaling factor is fixed at $0.6$. Furthermore, to train the generated CNN models, we have used Xavier weight initialization \cite{xavier} with the learning rate $0.001$ due to its effective utilization in the domain of DL. To enhance the training speed, we have incorporated batch normalization (BN) \cite{bn1} with a batch size of $32$, along with a $25\%$ dropout rate. The fitness calculation is conducted for each epoch throughout the evaluation procedure. The final CNN model architecture obtained from this proposed method is tested using the test dataset to evaluate its performance.
\begin{figure}[htbp]
\centerline
{\includegraphics[width=7.4cm, height=4.5cm]{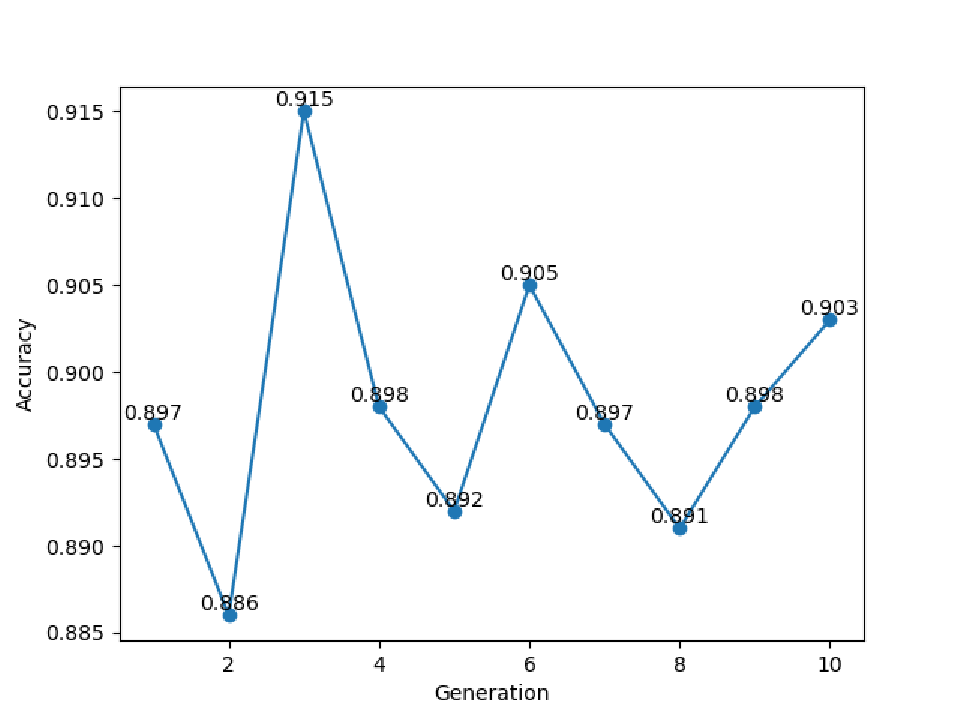}}
\caption{Generation wise accuracy plot for the proposed DE-based hyper-parameters selection approach.}
\label{fig:DE-wise-top}
\end{figure}
\par
\begin{figure}[htbp]
\centerline
{\includegraphics[height=1cm, width=7.4cm]{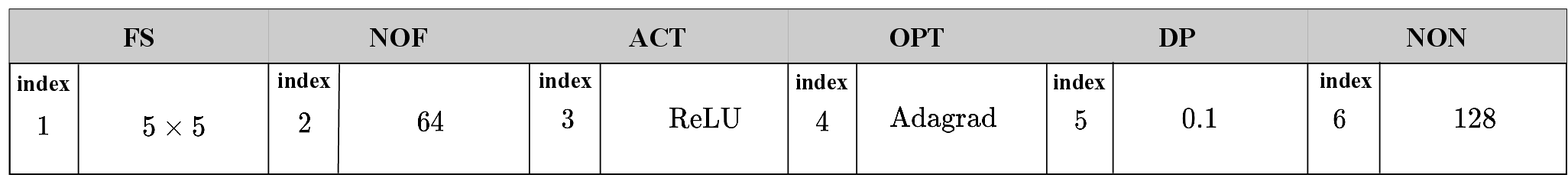}}
\caption{Hyper-parameters of the best CNN model (in terms of accuracy) obtained using DE-based hyper-parameters selection approach.}
\label{fig:GA-Hyper-wise-top}
\end{figure}
\begin{figure}[htbp]
\centerline
{\includegraphics[width=7.4cm, height=5.5cm]{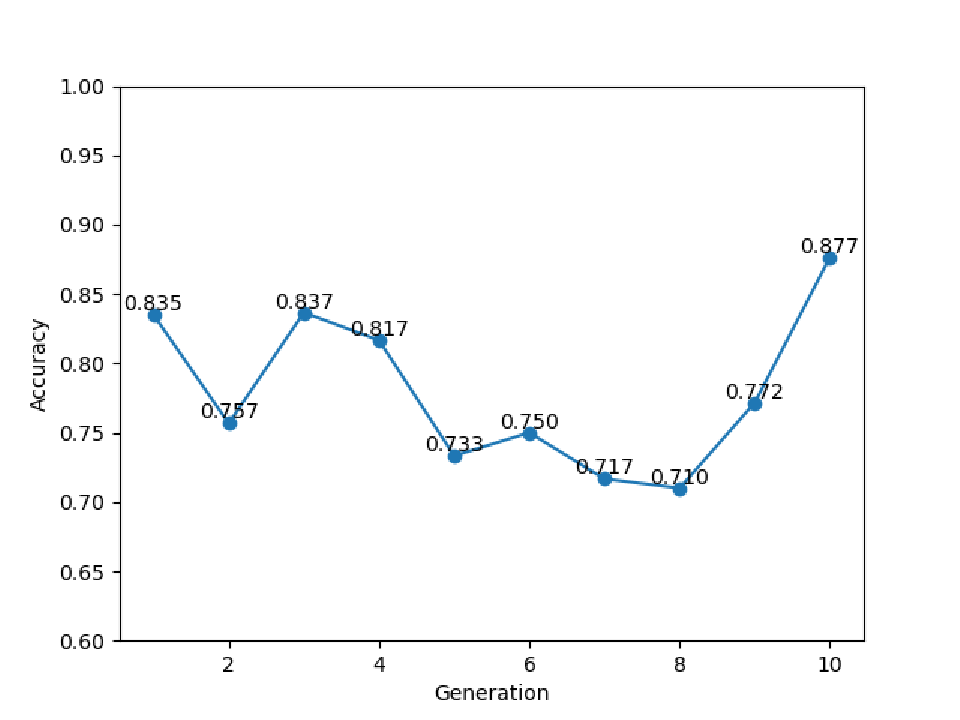}}
\caption{Generation wise accuracy plot for the proposed GA-based hyper-parameters selection approach.}
\label{fig:GA-wise-top}
\end{figure}
\begin{figure}[htbp]
\centerline
{\includegraphics[height=1cm, width=7.4cm]{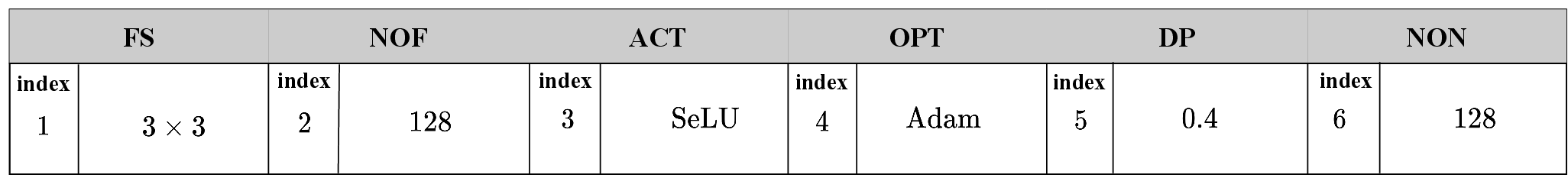}}
\caption{Hyper-parameters of the best CNN model (in terms of accuracy) obtained using GA-based hyper-parameters selection approach.}
\label{fig:-wise-top}
\end{figure}

In  Fig.\ref{fig:DE-wise-top}, the generation wise performance for the best networks of the proposed DE-based hyper-parameters selection approach are shown. The best networks for each generation indicate the best selection of hyper parameters belonging to the respective CNN networks. As shown in Fig.\ref{fig:DE-wise-top}, the highest test accuracy obtained is $0.915$ (i.e. $91.5\%$) for the generation number $3$. In Fig.\ref{fig:GA-Hyper-wise-top}, the hyper-parameters for the CNN model are presented for which the highest accuracy is obtained.  
The proposed approach is also compared with the GA-based hyper-parameter selection approach. In GA based approach, each chromosome is selected in each generation from the population size $15$. The generation-wise accuracy plot for the GA-based hyper-parameters selection approach is shown in Fig.\ref{fig:GA-wise-top}. From Fig.\ref{fig:GA-wise-top}, it can be observed that the highest accuracy obtained is $0.877$ (i.e. $87.7\%$) for the generation number $10$. 
In Fig.\ref{fig:-wise-top}, the hyper-parameters for the CNN model are presented for which the highest accuracy is obtained.  However, from both Fig.\ref{fig:DE-wise-top} and Fig.\ref{fig:GA-wise-top} it can be clearly observed that the performance (in terms of accuracy) of the DE-based hyper-parameter selection approach is better than the GA-based hyper-parameter selection approach. The performance of both DE and GA approaches are also compared with ResNet-50, Inception-V3, Xception, VGG-16, and VGG-19 models for the SCR task considering the test dataset. 

Table \ref{DATASET} presents the average precision, recall, F1-score, and test accuracy of all models considered, highlighting the superior performance of the proposed DE-based CNN model. While the ResNet-50 model also performs significantly better than other considered DCNN models, it is outshined by the proposed model.
\begin{table}[htbp]
\centering
\caption{Precision, Recall, F1-Score and Accuracy for all the considered models (averaged over 10 runs)}\label{DATASET}
\resizebox{1.0\linewidth}{!}{\begin{tabular}{c|c|c|c|c}
\hline
Models      & Precision & Recall & F1-Sore & Accuracy \\ \hline
ResNet-50   & \textbf{0.917}      & 0.907   & 0.908    & 0.908     \\ \hline
InceptionV3 & 0.892      & 0.887   & 0.884    & 0.886     \\ \hline
Xception    & 0.804      & 0.802   & 0.802    & 0.808     \\ \hline
VGG-16      & 0.828      & 0.820   & 0.819    & 0.823     \\ \hline
VGG-19      & 0.798      & 0.789   & 0.789    & 0.795     \\ \hline
GA-best     & 0.875      & 0.786   & 0.871    & 0.877     \\ \hline
DE-best     & \textbf{0.916}      & \textbf{0.914}  & \textbf{0.913}    & \textbf{0.915}     \\ \hline
\end{tabular}}
\end{table}

\begin{figure}[htbp]
\centerline
{\includegraphics[width=8cm, height=7cm]{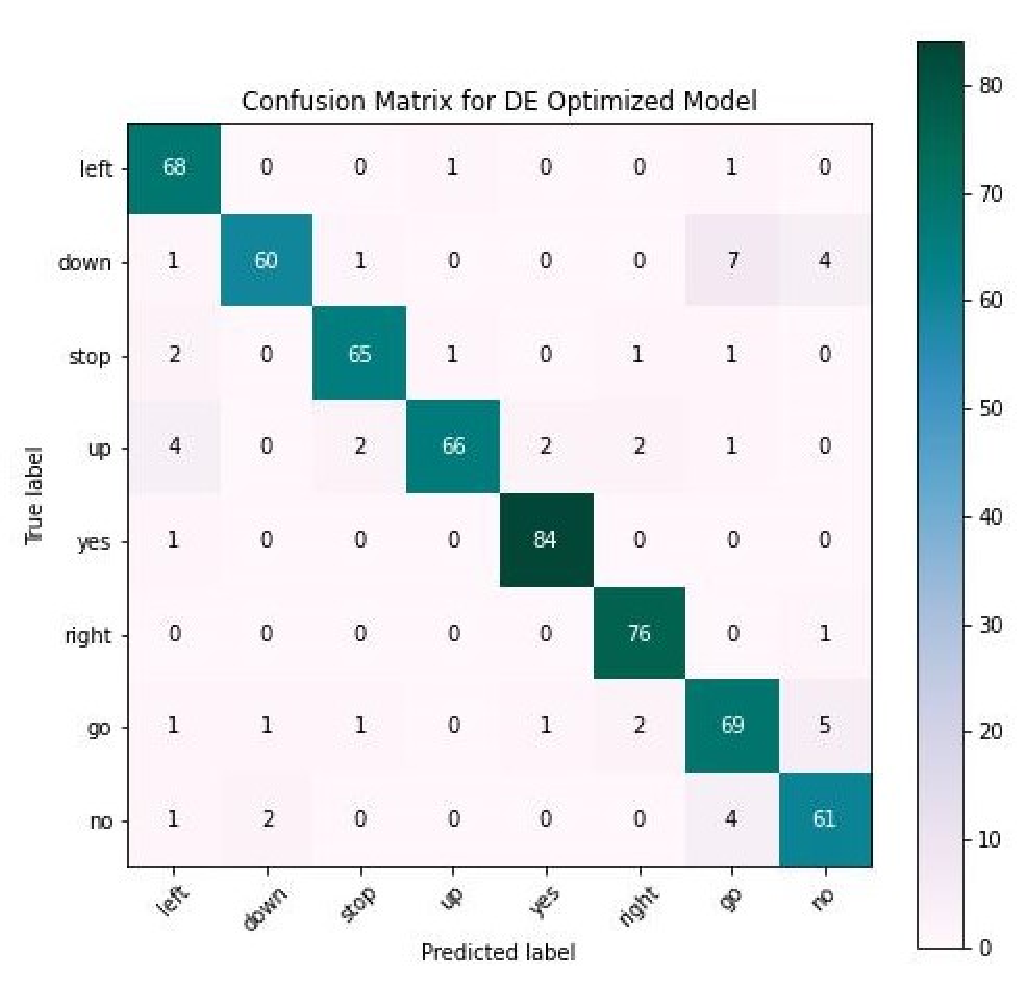}}
\caption{Confusion matrix of the best model obtained from the proposed approach.}
\label{fig:12DE-wise-top}
\end{figure}

However, from Table \ref{DATASET} it is clearly observed that the accuracy of the CNN model obtained from the proposed approach is higher than ResNet-50 model for the SCR task. Fig.\ref{fig:12DE-wise-top} shows the confusion matrix obtained from evaluating the model's class wise prediction accuracy for the DE approach on the test dataset.
From Fig.\ref{fig:12DE-wise-top}, it can be concluded that the CNN model obtained from the proposed approach has shown significant performance (in terms of accuracy) for all the considered classes.
\par

The superior performance of the DE-optimized CNN model is due to its effective exploration of the search space, utilizing parameter vector differences to exploit promising regions for optimal solutions, unlike genetic algorithms that may converge to local minima. The mutation operator prevents early convergence through random perturbations, while the crossover operator accelerates convergence by exchanging useful features.
\begin{table}[htbp]
\centering
\caption{Precision, Recall, F1-Score and Accuracy for all the considered GSC Dataset Speech Commands}\label{DATASET-121}
\small 
\resizebox{.8\linewidth}{!}
{\begin{tabular}{c|c|c|c}
\hline
Commands     & Precision & Recall & F1-Sore \\ \hline
left   & {0.871}      & 0.971   & 0.918     \\ \hline
down & 0.952      & 0.821   & 0.882     \\ \hline
stop   & 0.942     & 0.928   & 0.935   \\ \hline
up      & 0.970      & 0.857   & 0.910      \\ \hline
yes      & 0.965      & 0.988   & 0.976     \\ \hline
right     & 0.938      & 0.987   & 0.962   \\ \hline
go    & 0.831      & 0.862   & 0.846    \\ \hline
no     & {0.859}      & {0.897}  & {0.877}    \\ \hline
\end{tabular}}
\end{table}
The selection operator preserves the fittest individuals, enhancing the quality of solutions. Therefore, in Table \ref{DATASET-121} the class wise precision, recall, F1-score are also provided to show the performance of the obtained CNN model for all the considered speech commands of the GSC dataset. 

\section{\uppercase{Conclusion}}
\label{sec:conclusion}

This paper proposes an efficient Differential Evolution (DE)-based approach for selecting CNN hyper-parameters automatically, aiming to enhance Speech Command Recognition (SCR) tasks. Unlike tedious manual selection, DE, a global optimization algorithm, avoids local optima entrapments common in Grid Search, promoting more efficient promoting more efficient global optimum identification. Furthermore, evolutionary algorithms like DE inherently minimize user bias - when hyper-parameters are manually selected, they are often influenced by an individual's past experiences or preconceived notions, which can skew the optimization process. The proposed DE-based hyper-parameter selection approach outperformed the GA-based approach and other considered DCNN models in SCR tasks. The improved performance is attributed to DE's superior search space navigation and global maxima identification abilities. Unlike GA, DE requires fewer control parameters and has demonstrated robustness across various optimization problems. Additionally, the proposed approach surpassed other DCNN models. Future work may extend this approach to evolutionary algorithm-based speech feature selection for diverse speech-based applications.
\bibliographystyle{apalike}
{\small
\bibliography{example}}

\end{document}